\documentclass[aps, prd,letterpaper,showpacs,showkeys,twocolumn,floatfix,nofootinbib,superscriptaddress]{revtex4-1} 

\usepackage{amsmath,amssymb,amsfonts,graphicx}
\usepackage{bm}
\usepackage{slashed}
\usepackage{subfigure}
\usepackage[svgnames]{xcolor}

\newcommand{\vect}[1]{\boldsymbol{#1}}

\begin{document}

\title{Effects of Quark Spin Flip on the Collins Fragmentation Function in a Toy Model}

\preprint{ADP-12-21/T788}

\author{Hrayr~H.~Matevosyan}
\affiliation{CSSM and ARC Centre of Excellence for Particle Physics at the Tera-scale,\\ 
School of Chemistry and Physics, \\
University of Adelaide, Adelaide SA 5005, Australia
\\ http://www.physics.adelaide.edu.au/cssm
}

\author{Anthony~W.~Thomas}
\affiliation{CSSM and ARC Centre of Excellence for Particle Physics at the Tera-scale,\\ 
School of Chemistry and Physics, \\
University of Adelaide, Adelaide SA 5005, Australia
\\ http://www.physics.adelaide.edu.au/cssm
}

\author{Wolfgang Bentz}
\affiliation{Department of Physics, School of Science,\\  Tokai University, Hiratsuka-shi, Kanagawa 259-1292, Japan
\\ http://www.sp.u-tokai.ac.jp/
}

\begin{abstract}
 The recent extension of the NJL-jet model to hadronization of transversely polarized quarks allowed the study of the Collins fragmentation function. Both favored and unfavored Collins fragmentation functions were generated, the latter purely by multiple hadron emissions, with $1/2$ moments of opposite sign in the region of the light-cone momentum fraction $z$ accessible in current experiments. Hints of such behavior has been seen in the measurements in several experiments. Also, in the transverse momentum dependent (TMD) hadron emission probabilities, modulations of up to fourth order in sine of the polar angle were observed, while the Collins effect describes just the linear modulations. A crucial part of the extended model was the calculation of the quark spin flip probability after each hadron emission in the jet. Here we study the effects of this probability on the resulting unfavored and favored Collins functions by setting it as a constant and use a toy model for the elementary single hadron emission probabilities. The results of the Monte Carlo simulations showed that  preferential quark spin flip in the elementary hadron emission is needed to generate the favored and unfavored Collins functions with opposite sign $1/2$ moments. For the TMD hadron emission modulations, we showed that the model quark spin flip probabilities are a partial source of the higher rode modulations, while the other source is the Collins modulation of the remnant quark from the hadron emission recoil. 
\end{abstract}

\pacs{13.60.Hb,~13.60.Le,~13.87.Fh,~12.39.Ki}
\keywords{Collins fragmentation functions, TMDs, NJL-jet model, Monte Carlo simulations}

\date{\today}                                           

\maketitle

\section{Introduction}

 The recent interest in the transverse momentum structure of hadrons has prompted renewed efforts both in experimental and theoretical studies. Notably, experiments using semi-inclusive deep inelastic scattering (SIDIS) have advanced our understanding of the underlying partonic  structure of the nucleon in momentum space. For example, the naively time-reversal odd Sivers distribution function has been measured to be non-zero \cite{Airapetian:2004tw,Avakian:2003pk, Bradamante:2011xu}. In the factorization regime, the SIDIS cross-sections can be expressed as convolutions of parton distribution functions with elementary parton-probe scattering amplitudes calculable using perturbative QCD and parton fragmentation functions \cite{Collins:1985ue,Collins:1989gx}. The complete tree-level expressions at leading order have been presented in Ref~\cite{Mulders:1995dh}. Thus, to access the underling partonic structure of the nucleons, a precise knowledge of the fragmentation functions are needed. The particularly interesting case of the T-odd Collins fragmentation function $H_1^\perp$ has been widely explored both in theoretical models~\cite{Amrath:2005gv,Bacchetta:2007wc, Gamberg:2003eg,Artru:2010st} and experiments~\cite{Abe:2005zx,Seidl:2008xc,Airapetian:2004tw,Bradamante:2011xu,Aghasyan:2011ha,Aghasyan:2011gc}, with the first direct measurements of the Collins fragmentation mechanism being performed by the Belle collaboration using hadron pair production in $e^+ e^-$ collisions~\cite{Abe:2005zx,Seidl:2008xc}. Further, the experimental studies from HERMES, COMPASS and JLab are suggesting that the unfavored Collins functions have a similar size and an opposite sign to that of the favored ones~\cite{Airapetian:2004tw,Bradamante:2011xu,Aghasyan:2011ha,Aghasyan:2011gc}.
 
\begin{figure}[tbp]
\centering\includegraphics[width=1\columnwidth]{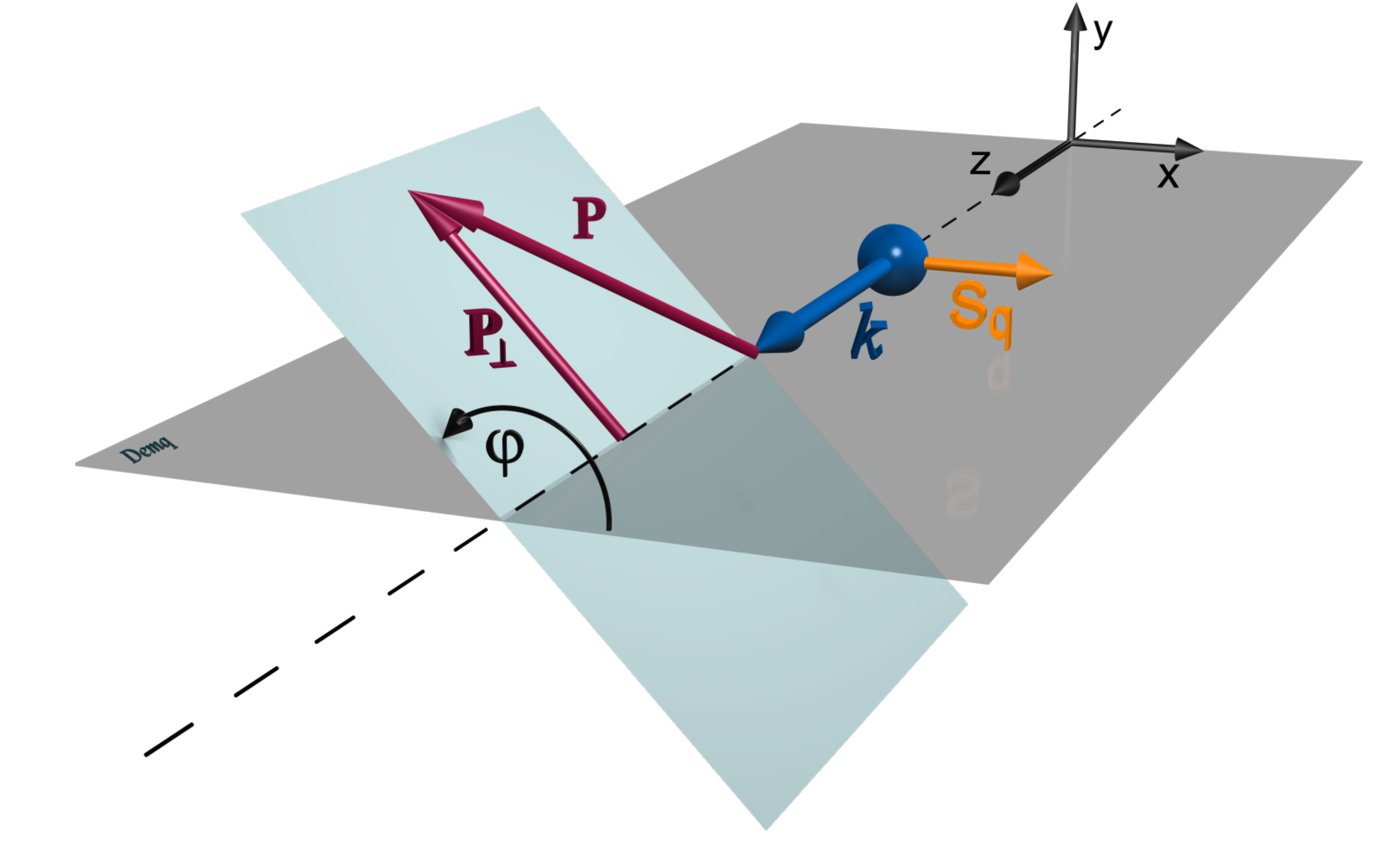}
\caption{Illustration of the three dimensional kinematics of transversely polarized quark fragmentation. 
The fragmenting quark's momentum $\vect{k}$ defines the $z$-axis with its transverse polarization spin vector $\vect{S}_q$ along $x$ axis. The emitted hadron has momentum $p$ with the transverse component $\vect{P_{\perp}}$ with respect to the $z$-axis. The polar angle of hadron's momentum $P$ with respect to the $z x$ plane is denoted by $\varphi$.}
\label{PLOT_POL_QUARK_3D}
\end{figure}

  The theoretical calculation of Collins function for pions and later also kaons within the spectator model of Refs.~\cite{Amrath:2005gv,Bacchetta:2007wc, Gamberg:2003eg} made predictions for the favored Collins function using the mechanism of interference of one-meson-loop amplitudes with the tree level amplitude. However, these model calculations could not make direct calculations of the unfavored Collins function, as only a single hadron emission was included. Recently, the Nambu--Jona-Lasinio (NJL) jet model of Refs.~\cite{Ito:2009zc, Matevosyan:2010hh,Matevosyan:2011ey, Matevosyan:2011vj,Casey:2012ux} has been extended to calculate the Collins fragmentation functions, both favored and unfavored, using Monte Carlo (MC) simulations in the quark-jet hadronization mechanism~\cite{Matevosyan:2012ga}. Here the transverse polarization of the fragmenting quark has been introduced, and the light cone spinors were used to calculate the quark spin flip probabilities in each hadron emission step. In the hadronization of a transversely polarized quark, the Collins effect describes a modulation of the unpolarized hadron fragmentation function with a Collins fragmentation function term that is proportional to the sine of the angle $\varphi$ of the hadron's transverse momentum and the quark's spin in the $\gamma^*N$ frame, as depicted in Fig.~\ref{PLOT_POL_QUARK_3D}.
The corresponding expression using the "Trento Convention" \cite{Bacchetta:2004jz}, with polarized quark $q$  
carrying momentum $\vect{k}$ and spin $\vect{S_q}$ fragmenting to unpolarized hadron $h$ of mass $m_h$, carrying the light-cone momentum fraction $z$ and transverse momentum $\vect{P}_\perp$ with respect to quark's momentum $\vect{k}$ can be written as 
\begin{align}
\label{EQ_Dqh_SIN}
D_{h/q^{\uparrow}} (z,P_\perp^2,\varphi) &= D_1^{h/q}(z,P_\perp^2)\\ \nonumber
 &- H_1^{\perp h/q}(z, P_\perp^2) \frac{ P_\perp S_q}{z m_h} \sin(\varphi),
\end{align}
where the unpolarized fragmentation function is denoted $D_1^{h/q}(z,P_\perp^2)$ and $H_1^{\perp h/q}(z, P_\perp^2)$ is the  Collins function. When integrated over $P_\perp^2$, the polarized fragmentation function can be expressed in terms of the integrated unpolarized fragmentation function $D_1^{h/q}(z)$ and the $1/2$ moment of the Collins function $H_{1 (h/q)}^{\perp  (1/2)}(z)$
 \begin{align}
\label{EQ_Nqh_Phi}
D_{h/q^{\uparrow}}(z,\varphi) &\equiv \int_0^{\infty} d P_\perp^2\ D_{h/q^{\uparrow}}(z,P_\perp^2,\varphi) \\ \nonumber
=  \frac{1}{2\pi}&\left[D_1^{h/q}(z)\ - 2H_{1 (h/q)}^{\perp  (1/2)}(z) S_q \sin(\varphi) \right]dz\ d \varphi,
\end{align}
where
\begin{align}
\label{EQ_D1}
D_{1}^{h/q}(z) &\equiv \pi \int_0^{\infty} d P_\perp^2\ D_1^{h/q}(z, P_\perp^2),\\
\label{EQ_H12}
H_{1 (h/q)}^{\perp  (1/2)}(z) &\equiv \pi \int_0^{\infty} d P_\perp^2 \frac{P_\perp}{2z m_h}  H_1^{\perp h/q}(z, P_\perp^2).
\end{align}

 In the Monte Carlo simulations of Ref.~\cite{Matevosyan:2012ga}, we used as input the elementary unpolarized and Collins fragmentation functions for single hadron production, calculated within NJL model, where we used the mechanism of Ref.~\cite{Bacchetta:2007wc} for the elementary Collins function. The full polarized fragmentation function was then calculated in MC simulations with multiple hadron emission. The resulting polarized fragmentations, when integrated over the hadron transverse momentum squared ($P_\perp^2$), exhibited the Collins modulation both for favored and unfavored hadron production. Here the unfavored Collins function's $1/2$ moment is comparable in size but opposite in sign to that for the favored fragmentation. For the transverse momentum dependent (TMD) polarized fragmentation functions, at fixed values for the hadron's light cone momentum fraction $z$ and $P_\perp^2$, exhibited modulation with a fourth order polynomial in $\sin(\varphi)$, even in simulations with only two produced hadrons. 
 
  In this work we have two goals, focusing on pion fragmentations from light quarks only. First, we examine the dependence of the sign and the magnitude of the $1/2$ moment of unfavored fragmentation function on the quark spin flip probability used in the NJL-jet model MC simulations. Second, we examine the origins of the higher order Collins modulations in the TMD polarized fragmentation functions.
    
 \section{NJL-jet Model with transversely polarized quark}
 \label{SEC_NJL-JET-SPIN}
 
\begin{figure}[htbp]
\centering\includegraphics[width=1\columnwidth]{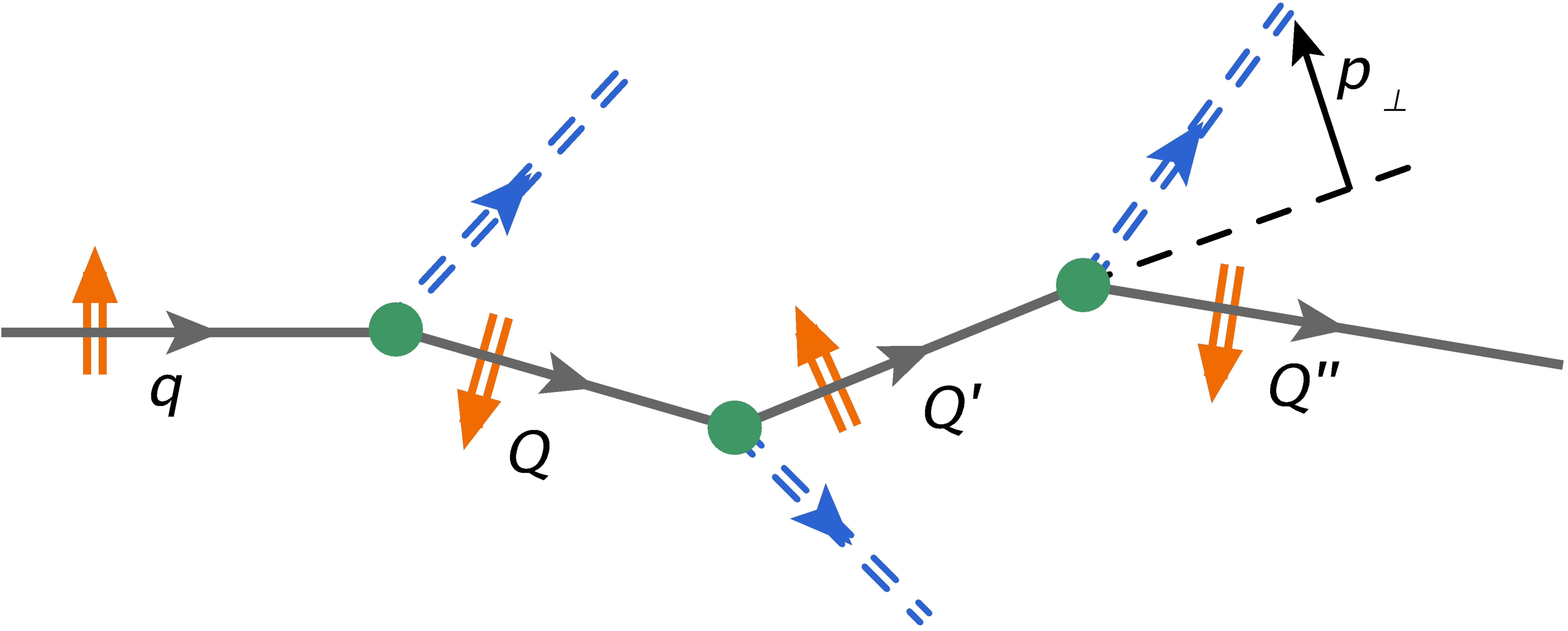}
\caption{NJL-jet model including transverse momentum and quark polarization transfer. Here the orange double-lined arrows schematically indicate the spin direction of the quark in the decay chain.}
\label{PLOT_NJL-JET_TMD}
\end{figure}

 Schematically, the NJL-jet model with the transversely polarized quark is depicted in Fig.~\ref{PLOT_NJL-JET_TMD}. In MC simulations of Ref.~\cite{Matevosyan:2012ga}, in each hadronization step, we used the elementary polarized fragmentations to sample the momentum of the production hadron, $z_1$, $p_\perp$ and $\varphi_1$, with respect to the fragmenting quark in that step (here $\varphi_1$ is defined with respect to the spin of the fragmenting quark). Using the momentum we calculated and recorded the hadron's momentum, $z$, $P_\perp$ and $\varphi$, with respect to the initial fragmenting quark. Further, we also determined the momentum of the remnant quark using momentum conservation, while the probability of the remnant quark's spin flip was sampled using the spin non-flip and flip probabilities $|a_1|^2$ and $|a_{-1}|^2$. These probabilities were calculated using Dirac spinors of the transversely polarized quark, that in turn can be expressed in terms of the Lepage-Brodsky spinors in helicity basis \cite{Lepage:1980fj, Brodsky:1997de}. Explicit these probabilities are proportional to
\begin{align}
\label{EQ_SPIN_FLIP}
| a_1 |^2\sim {l_x^2},\ | a_{-1} |^2 \sim {l_y^2+(M_2-(1-z)M_1)^2},
\end{align}
where $M_1$ and $M_2$ are the masses of the fragmenting and the remnant quarks, and $\vect{l}_\perp \equiv (l_x,l_y)=-\vect{p_\perp}$ is the transverse momentum of the remnant quark with respect to the momentum direction of the fragmenting quark.

 In this work, to explore the role of the spin-flip probability, we perform MC simulations with various fixed values of this probability. This allows us to clearly demonstrate the role of the quark spin flip during the hadronization process on the resulting fragmentation functions. Also, using constant for this probability instead of the one with the dynamical factors of Eq.~(\ref{EQ_SPIN_FLIP}) will help to disentangle the sources of the higher order modulation effects in $\sin(\varphi)$ observed in TMD fragmentations. 

For input to the Monte Carlo framework, the elementary probability of a transversely polarized quark $q$ to emit a single hadron $h$ carrying light-cone momentum fraction $z$, transverse momentum $\vect{p}_\perp$ (with magnitude $p_\perp$ and polar angle $\varphi$ with respect to the emitting quark's spin), can be written as
\begin{align}
\label{EQ_ELEM_Dqh_SIN}
\nonumber
d_{h/q^{\uparrow}} (z,p_\perp^2,\varphi)= d_1^{h/q}&(z,p_\perp^2)\\
&-\widetilde{H}_1^{\perp h/q}(z, p_\perp^2) \frac{ p_\perp S_q}{z m_h} \sin(\varphi),
\end{align}
where $ d^{h/q}(z,p_\perp^2)$ is the elementary unpolarized fragmentation function and $\widetilde{H}_1^{\perp h/q}(z, p_\perp^2)$ is the elementary Collins fragmentation function. The unpolarized fragmentation function has been calculated within the NJL model in tree level approximation~\cite{Matevosyan:2011vj}. The corresponding expression for an unpolarized quark emitting a pseudoscalar meson can be written as
\begin{align}
\label{EQ_QUARK_FRAG_TMD}
d_{1}^{h/q}(z,p_\perp^2)
&= G_{hQ}(p_\perp^2) \frac{C_q^h }{16\pi^{3}}\, g_{hqQ}^{2}\, z\ \\\nonumber
&\times \frac{p_{\perp}^{2} + \left[(z-1)M_{1}+M_{2}\right]^{2}} 
{\left[p_{\perp}^{2}+z(z-1)M_{1}^{2}+zM_{2}^{2}+(1-z)m_{h}^{2}\right]^{2}},
\end{align}
where the subscripts on the constituent masses, $M_1$ and $M_2$, denote the flavors of the fragmenting and the remnant quarks respectively, that are also indicated by the subscripts $q$ and $Q$. $C_q^h$ and $g_{hqQ}$ denote the isospin factor and quark-meson coupling constant, that have been determined  within the NJL model~\cite{Matevosyan:2010hh,Matevosyan:2011vj}, while $G_{hQ}(p_\perp^2)$ is the dipole regulator for the transverse momentum dependance.  The details of the derivations and the model parameters can be found in Ref.~\cite{Matevosyan:2011vj}

\begin{figure}[phtb]
\centering 
\includegraphics[width=0.9\columnwidth]{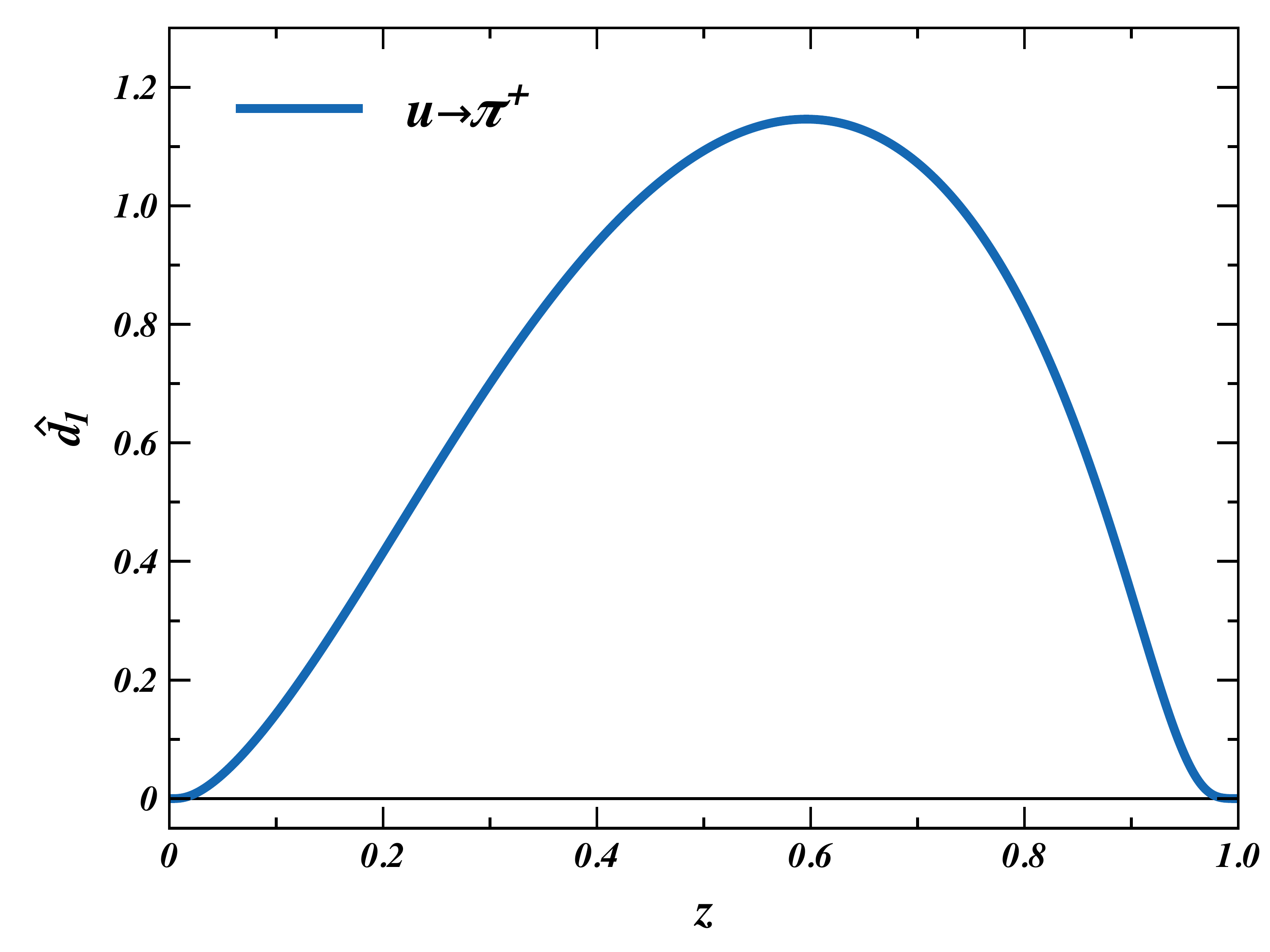}
\caption{Elementary renormalized unpolarized fragmentation function, $\hat{d}_1^{\pi^+/u}$ used as input in the Monte Carlo simulations.}
\label{PLOT_ELEM_FRAG}
\end{figure}

The model elementary Collins function within the NJL model has been also calculated in Ref.~\cite{Matevosyan:2012ga}, using the mechanism of interference between tree level amplitude and amplitude with gauge link coupling for single hadron emission, outlined in Refs.~\cite{Bacchetta:2007wc, Gamberg:2003eg}. Here we are interested in exploring the general properties of the quark-jet hadronization mechanism, thus the details of the particular models are unimportant. To further simplify the model and help reduce the required number of the Monte Carlo events that allow for a reliable extraction of the higher order modulations by enhancing the azimuthal modulation signal with respect to the statistical fluctuations, we employ a toy model for the elementary Collins function taking it promotional to the unpolarized one, with a coefficient chosen such that the polarized probability is still positive for all values of the corresponding arguments
\begin{align}
\label{EQ_MC_DRV_TOY}
d_{h/q^{\uparrow}}^{(toy)} (z,p_\perp^2)=  d_1^{h/q}(z,p_\perp^2)(1-0.9 \sin{\varphi}),
\end{align}
where we assume that  $\widetilde{H}_{1 }^{\perp h/q}(z,p_\perp^2)\frac{p_\perp}{m_h z} = 0.9\ d_1^{h/q}(z,p_\perp^2)$ and $S_q=1$.  Within the NJL-jet model, the elementary splitting functions are renormalized such that quark's total probability of emitting a hadron in each step is one: $\sum_h \int dz\ dp_\perp^2/2\ d \varphi\ \hat{d}_{h/q^{\uparrow}} (z,p_\perp^2,\varphi)=1$, the sum is over all the hadrons, while a quark of given flavor can emit directly. These renormalized splittings will be used in the next two sections as input to the Monte Carlo simulations of the quark-jet hadronization process. The renormalized elementary unpolarized fragmentation function, integrated over $p_\perp^2$, for fragmentation of a $u$ quark to $\pi^+$ is depicted in Fig.~\ref{PLOT_ELEM_FRAG}. All the other elementary fragmentation functions from light quarks ($u$ and $d$) to pions are related to $d_1^{\pi^+/u}$ either by a simple isospin factor or isospin symmetry.

\section{The Quark Spin Flip Probability Effects on $1/2$ Moments of the Collins Fragmentation Functions }
\label{SEC_TOY_MODEL}

\begin{figure}[phtb]
\centering 
\subfigure[] {
\includegraphics[width=0.95\columnwidth]{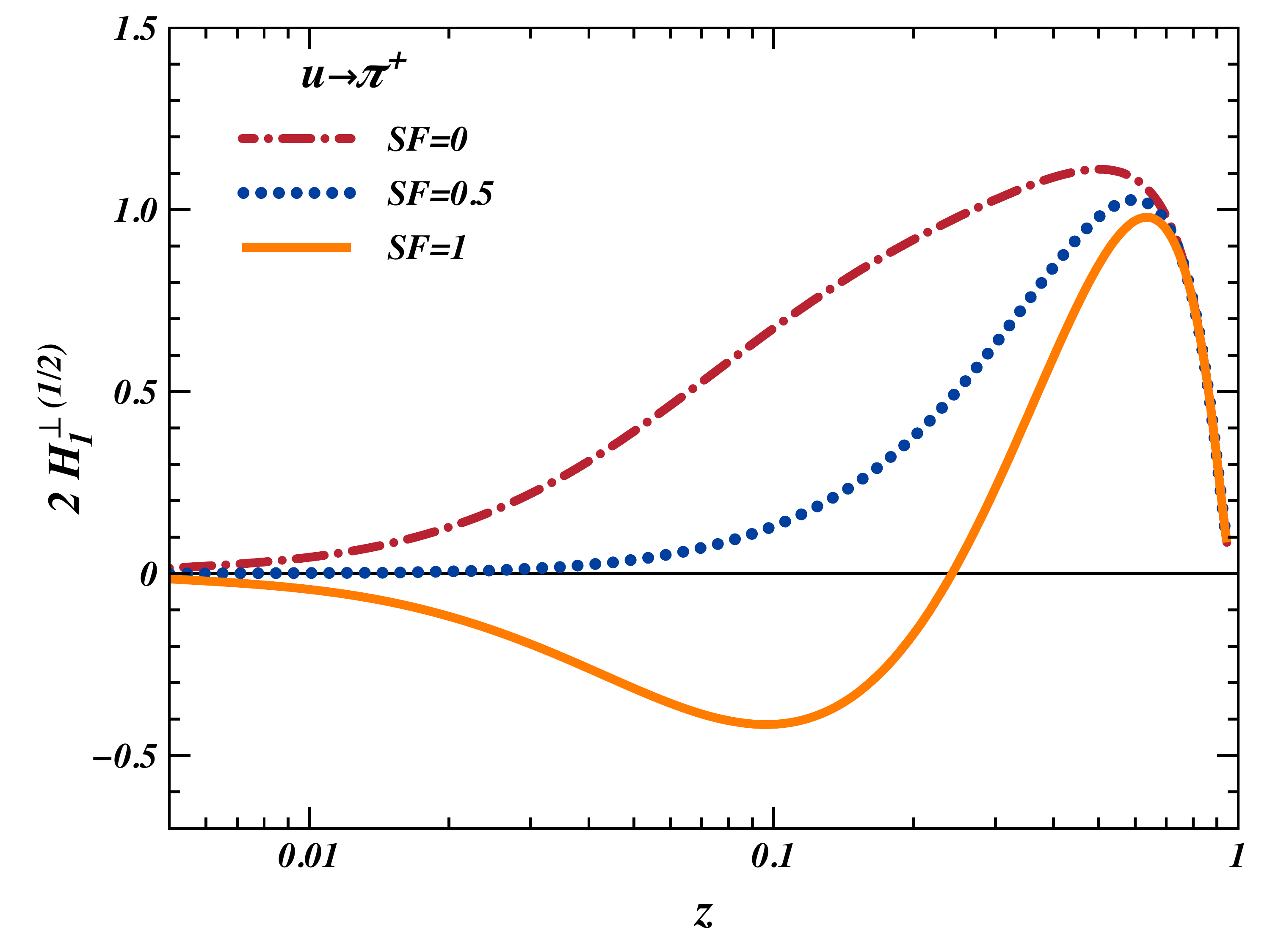}}
\hspace{0.cm} 
\subfigure[] {
\includegraphics[width=0.95\columnwidth]{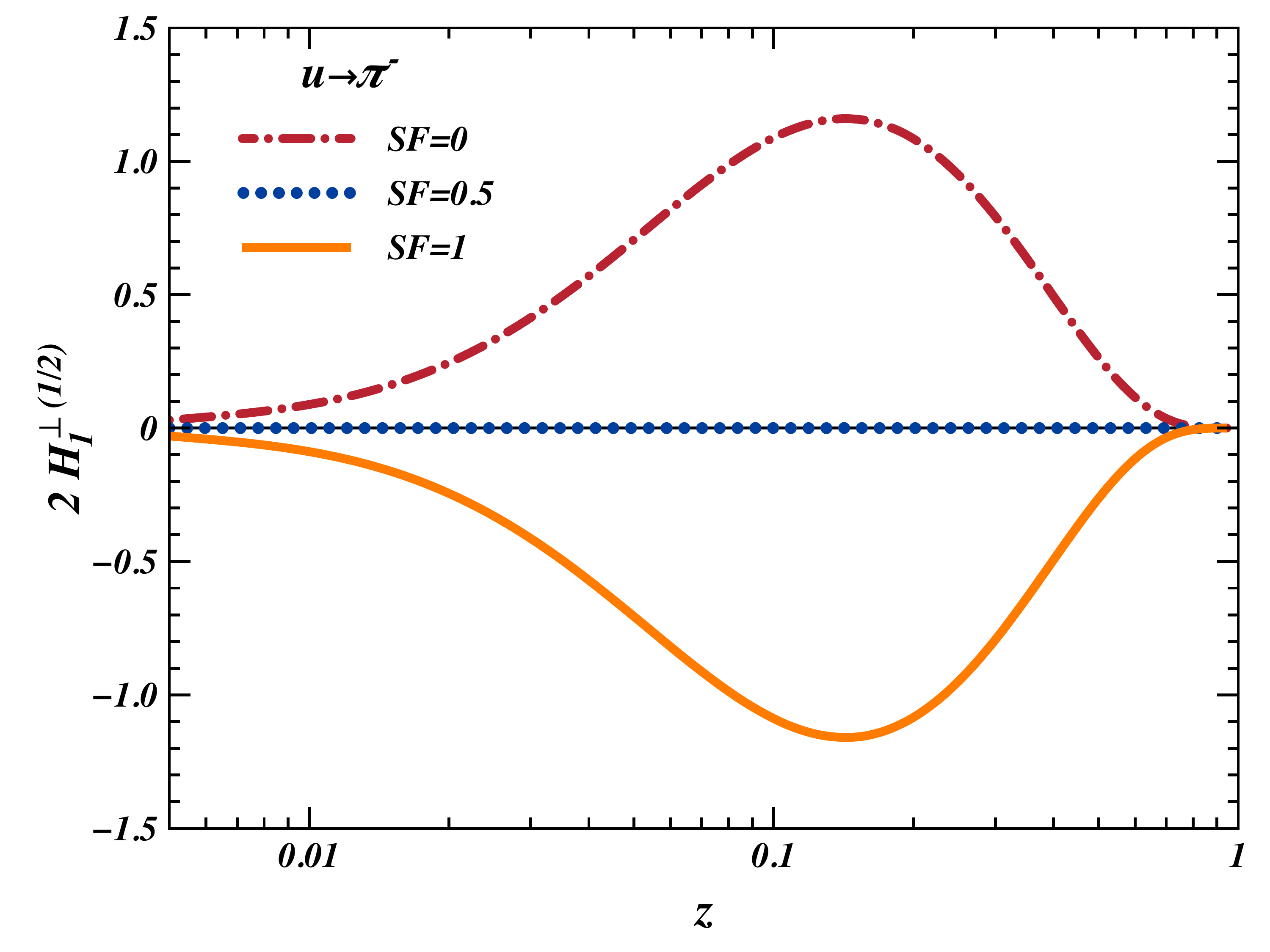}
}
\caption{Fitted values for twice the Collins function $1/2$ moment, $2 H_1^{\perp (1/2)}$ for $u\to\pi^+$ (a)  and $u\to\pi^-$ (b) as a function of $z$ from Monte Carlo simulations with three different values of the remnant quark spin flip probability $SF$ in each elementary emission, where the number of emitted hadrons in each quark decay chain is fixed to $N_{Links}=2$.}
\label{PLOT_H12_PI}
\end{figure}

  In this section we explore the properties of the $1/2$ moments of the Collins functions, Eq.~(\ref{EQ_H12}), for pions emitted in the hadronization of an up quark as a function of the quark spin flip probability $SF$ in each hadron emission step.  We perform Monte Carlo simulations to extract the polarized quark fragmentation functions $D_{h/q^{\uparrow}}$ as functions of  $z$, $P_\perp^2$ and $\varphi$ within the NJL-jet framework, as described in Ref.~\cite{Matevosyan:2012ga}. In this section, we limit the number of emitted hadrons in each quark decay chain to $N_{Links}=2$. This allows us to cleanly see the effects of the quark spin flip on the emission of the second hadron in the decay chain, where the unfavored fragmentation functions are first generated. 

  Next, to extract the integrated unpolarized fragmentation function~(\ref{EQ_D1}) and the $1/2$ moment of the Collins fragmentation functions~(\ref{EQ_H12}), we use the relation to the $P_\perp^2$ integrated transversely polarized fragmentation written in Eq.~(\ref{EQ_Nqh_Phi}). Thus, we perform a minimum-$\chi^2$ fit  to $D_{h/q^{\uparrow}}(z,\varphi)$ using a form $F(c_0,c_1) \equiv c_0+c_1 \sin(\varphi)$ for fixed values of $z$, identifying the fitted coefficients with the unpolarized and Collins terms in Eq.~(\ref{EQ_Nqh_Phi}). The resulting fits for the $P_\perp^2$ integrated pion polarized fragmentation functions with the functional form of $F(c_0,c_1)$ perfectly describe the results of the MC simulations, with average $\chi^2$ per number of degrees of freedom (NDF) very close to unity. The unpolarized fragmentations are not affected by the quark spin flip probability $SF$ and have been studied in detail in our previous work~\cite{Matevosyan:2011ey,Matevosyan:2011vj}. Thus we only present here the results for the Collins functions.

 The results from fits for $H_1^{\perp (1/2)}(z)$ for fragmentations of a $u$ quark to $\pi^+$ and $\pi^-$ are depicted in Fig.~\ref{PLOT_H12_PI}, where each curve depicts the results for a different value of $SF$.  The results for $\pi^+$ in the subplot (a) show that, if $SF\leq 0.5$, then $H_1^{\perp (1/2)}$ is always positive, crossing zero and becoming negative at small values of $z$ for $SF>0.5$. For unfavored $\pi^-$ fragmentation, the effects of $SF$ are more dramatic: $H_1^{\perp (1/2)}$ is positive (same sign as that for $\pi^+$) for $SF<0.5$ and vanishes for $SF=0.5$. Thus, in order to generate a non-zero unfavored Collins fragmentation function, the quark spin non-flip and flip must be unequally probable. Finally, for $SF>0.5$ the $H_1^{\perp (1/2)}$ is negative, at $SF=1$ becoming exactly the negative of the results for $SF=0$. This is easy to understand, as the unfavored functions are only generated after the initial hadron emission, so we can consider only two hadron emission for simplicity. In case where $SF=1$ (flip of the quark's spin after each emission), the angle $\varphi$ in Eq.~(\ref{EQ_ELEM_Dqh_SIN}) that is measured with respect to the spin of the initial quark acquires an additional phase $\pi$ compared to the simulations with $SF=0$ (no quark spin flip) since the modulation in each hadron emission step is generated with respect to the azimuthal angle of the currently fragmenting quark, resulting in the Collins term having the opposite sign. 

 Thus, the representative features of the full model results in Ref.~\cite{Matevosyan:2012ga}, where the $1/2$ moments behave similarly to the scenario where $SF>0.5$, qualitatively do not depend on the details of the particular input unpolarized and Collins fragmentation functions, but rather represent the characteristics of the quark-jet model with preferential spin flip probability ($|a_{-1}|^2>|a_1|^2$ in Eq.~(\ref{EQ_SPIN_FLIP})).
 
\section{Higher Order Modulations in TMD Collins Function}
\label{SEC_TMD}

\begin{figure}[ptbh]
\centering 
\subfigure[] {
\includegraphics[width=0.95\columnwidth]{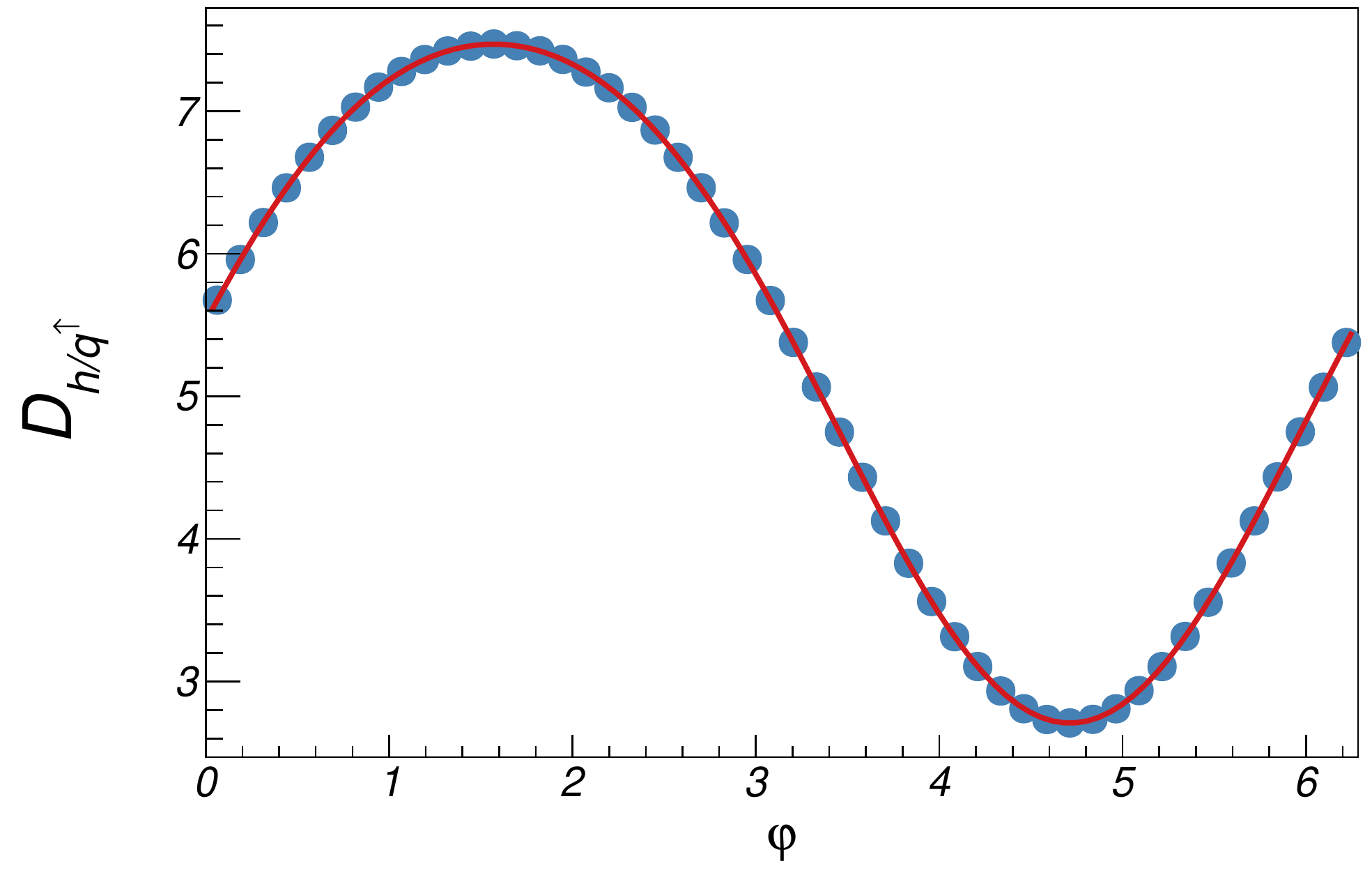}
}
\vspace{0.1cm} 
\subfigure[] {
\includegraphics[width=0.95\columnwidth]{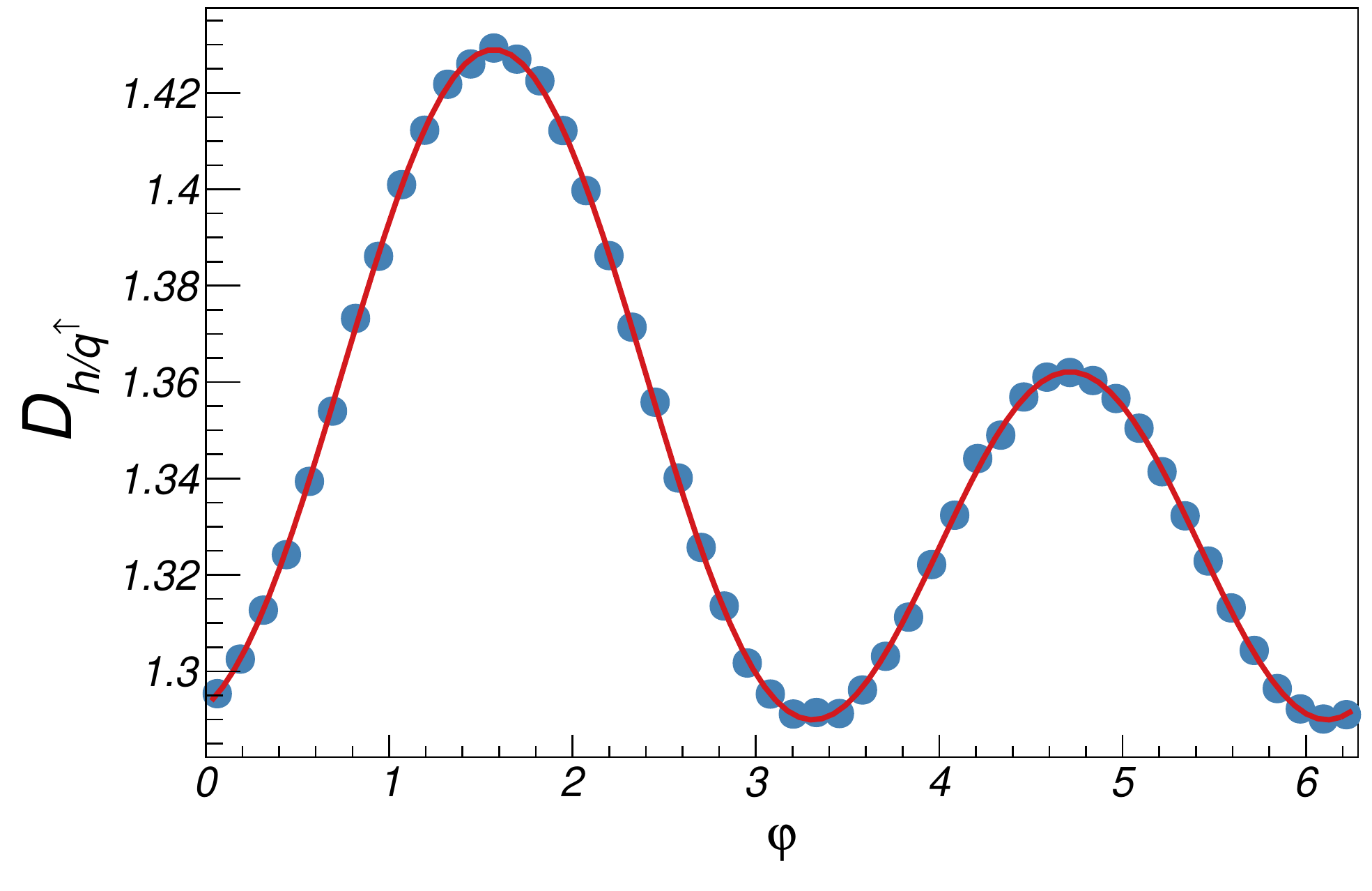}
}
\caption{The histograms (blue dots) for the polarized number density $D_{\pi^+/u^\uparrow}(z,P_\perp^2, \varphi)$ for fixed values of $P_\perp^2=0.04~\mathrm{GeV}^2$ (a) and $P_\perp^2=0.25~\mathrm{GeV}^2$ (b) with $z=0.3$ for both cases,  as a function of the azimuthal angle $\varphi$, from Monte Carlo simulations  using the toy model with a fixed number of hadrons emitted in each decay chain, $N_{Links}=2$. The minimum-$\chi^2$ fits with a second order polynomial in $\sin (\varphi)$  (red lines) are also depicted.}
\label{PLOT_COL_SLICE}
\end{figure}

 In the previous section we showed that setting quark spin flip probability $SF=1$ can be considered as the asymptotic solution for generating the largest value of the unfavored Collins function that mimics the behavior of our model calculations in Ref.~\cite{Matevosyan:2012ga}. Thus in this section we will concentrate on simulations with $SF=1$, that will allow us to easily extract the Collins modulations from the unpolarized fragmentation functions.
 
  In this section we consider the full transverse momentum dependent fragmentation functions. As mentioned in Ref.~\cite{Matevosyan:2012ga}, the polarized fragmentation function can no longer be successfully described with a first order polynomial in $\sin(\varphi)$ for all values of $z$ and $P_\perp^2$, even for MC results with just two hadrons emitted, $N_{Links}=2$. The plots in Fig.~\ref{PLOT_COL_SLICE} depict the MC result $D_{\pi^+/u^\uparrow}(z,P_\perp^2, \varphi)$ (in blue dots) for a fixed value of $z=0.3$ and $P_\perp^2=0.04~\mathrm{GeV}^2$ (a), and $P_\perp^2=0.25~\mathrm{GeV}^2$ (b). The minimum-$\chi^2$ fits with a second order polynomial in $\sin (\varphi)$ are depicted as red lines. 
 
\begin{figure}[phtb]
\centering 
\subfigure[] {
\includegraphics[width=0.95\columnwidth]{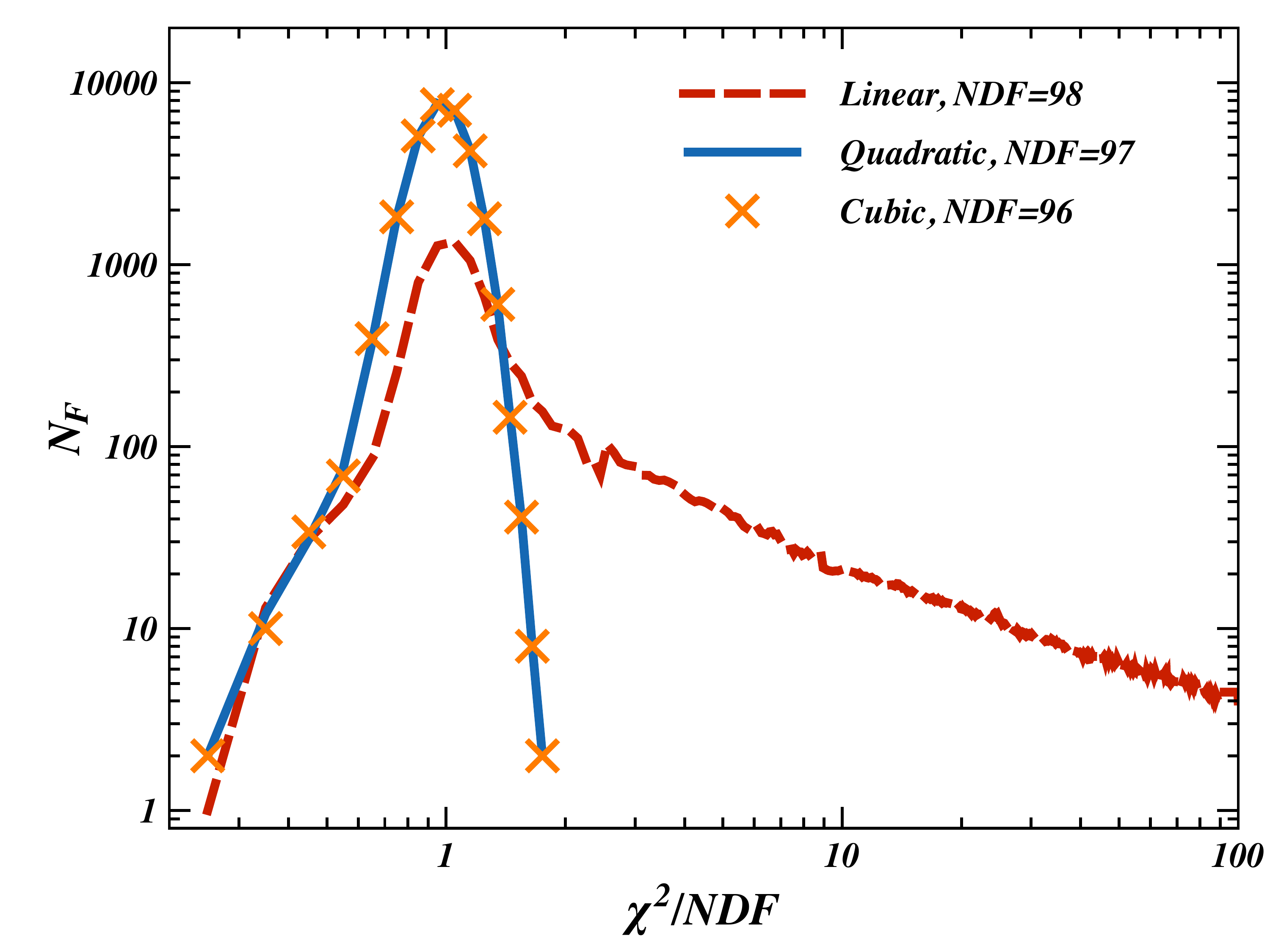}}
\hspace{0.cm} 
\subfigure[] {
\includegraphics[width=0.95\columnwidth]{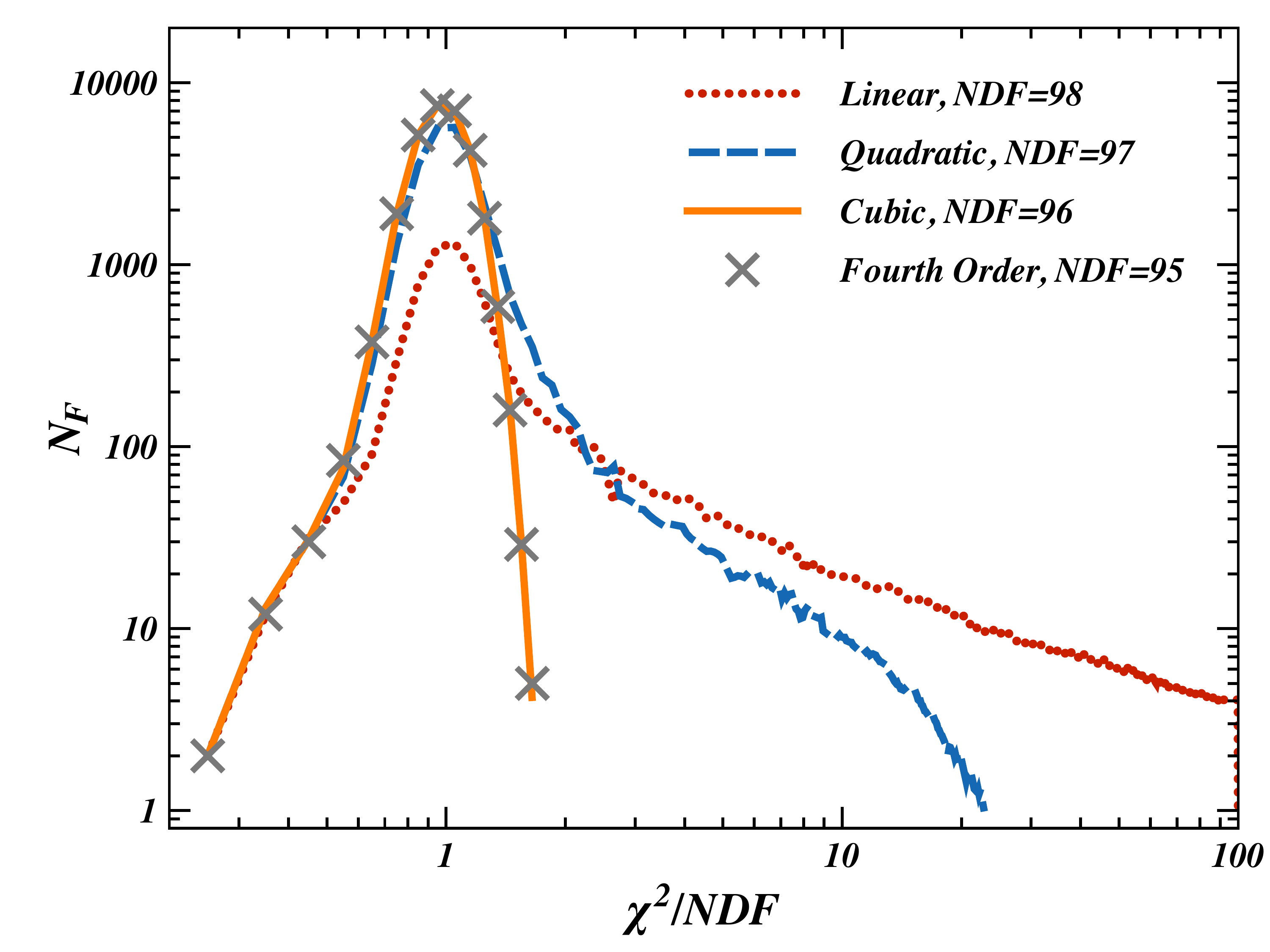}
}
\vspace{0.cm} 
\subfigure[] {
\includegraphics[width=0.95\columnwidth]{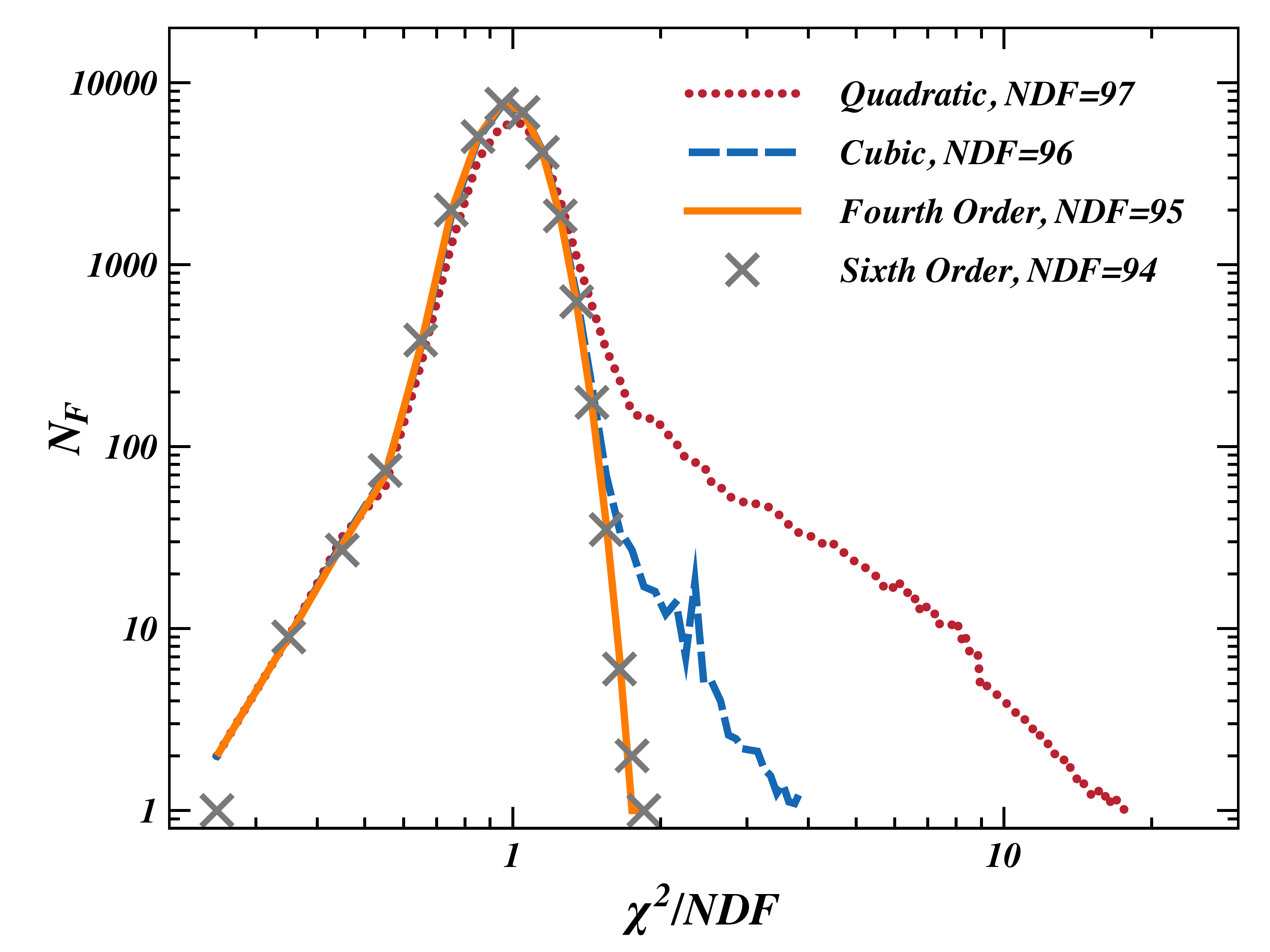}
}
\caption{The histograms of the values $N_F$ of $\chi^2$/NDF for all the hadrons produced by a $u$ quark with polynomials of different orders in $\sin(\varphi)$ for MC simulations for values of $N_{Links}$ of $2$ (a), $3$ (b) and $6$ (c).}
\label{PLOT_DPOL_FITS}
\end{figure}

Here we analyze the dependence of the order of these modulations on the number of hadrons produced, $N_{Links}$. For that, we consider minimum-$\chi^2$ fits of the corresponding polarized fragmentation functions with polynomials in $\sin(\varphi)$ of various orders, for fixed values of $z$ and $P_\perp^2$. To quantify the quality of  the fits, we use as a measure the $\chi^2/NDF$, that should be close to $1$ for a good fit. Our MC simulations are performed with $100$ discretization points for both $z$ and $P_\perp^2$  in the interval $[0,1)$ and $100$ discretization points for $\varphi$ in the interval $[0,2\pi)$. Thus, the number of the "data points" to fit, for all distinct values of $z$ and $P_\perp^2$, as well as all the possible pion fragmentations of $u$ quark (to $3$ types of pions) are $N_F=3\times10^4$. The histograms in Fig.~\ref{PLOT_DPOL_FITS} depict the results for these fits for simulations with $N_{Links}=2$ (a), $N_{Links}=3$ (b) and $N_{Links}=6$ (c). Here the horizontal axis denotes the $\chi^2/NDF$ and the vertical axis is the number of the fits that yield this value, while the different lines represent the results for fits with different order polynomials in $\sin(\varphi)$. The results for $N_{Links}=2$, depicted in Fig.~\ref{PLOT_DPOL_FITS}(a), show that the linear form of Eq.~(\ref{EQ_Dqh_SIN}) is inadequate to describe the polarized fragmentations for a significant number of fixed values of $z$ and $P_\perp^2$. The quadratic form, on the other hand, provides an excellent description of the simulations, and it can be confirmed by the histograms that cubic fits don't improve the resulting values of $chi^2/NDF$. Similarly, for the simulation with $N_{Links}=3$ of Fig.~\ref{PLOT_DPOL_FITS}(b), the cubic polynomial yields excellent fits, while the lower order polynomials are inadequate to fully describe the simulations. Finally, for the simulations with $N_{Links}=6$, depicted in Fig.~\ref{PLOT_DPOL_FITS}(c), at least a fourth order polynomial is needed for good fits, though there is a marginal improvement with fits using sixth order polynomials. 

 A simple explanation of these modulations within the quark-jet picture can be easily given. Here, after the initial hadron emission, the hadron emission probability is modulated according to the elementary polarized fragmentation of Eq.~(\ref{PLOT_ELEM_FRAG}). Thus, the remnant quark, having transverse recoil momentum with respect to the initial quark that is equal to minus that of the emitted hadron, also acquires opposite signed sine modulation in the distribution of its transverse momentum.  In the next hadron emission step, the emitted hadron probability is again modulated according to the elementary polarized fragmentation of Eq.~(\ref{PLOT_ELEM_FRAG}), but now with respect to the currently emitting quark, that itself has a sine modulation in its polar angle distribution. Thus, the produced hadron probability acquires a second order sine modulation in $\varphi$ with respect to the initial quark, as does the remnant quark in this step, and so on. Thus, in the simulations with $N_{Links}$ produced hadrons, the resulting polarized fragmentation functions should be modulated with a polynomial of order $N_{Links}$, as we have seen from the results of Figs.~\ref{PLOT_DPOL_FITS} (a) and (b), while the simulations with $N_{Links}=1$ yield the trivial linear modulation. We have to also note from our earlier discussions of the mechanism for this quark recoil traverse momentum induced modulations, that the hadrons emitted in $n$-th step will only have a emission probability modulation of the order $n$ in $\sin(\varphi)$. Also, we have shown in our previous work, that the remnant quark looses almost all of its light-cone momenta within just a few hadron emission steps~\cite{Matevosyan:2011ey}, so that the hadrons produced with higher and higher orders of modulations will have smaller and smaller average values of $z$. This explains the fact that the linear form gives reasonably good fits for significant number of slices even in the simulations with large number of $N_{Links}$. Moreover, the results in Fig.~\ref{PLOT_DPOL_FITS}(c) for $N_{Links}=6$ also can be explained using this argument, as our simulations, with only $100$ uniform discretization points in $z$, do not probe in detail the very low $z$ region, where the sixth order polynomial modulations will be significant. Thus the corresponding results can be well described with just fourth order polynomials. Also, the size of the statistical errors and relatively small number of the discretization points in $\varphi$ would affect the sensitivity to this higher order terms in the fits. 

 In the full model calculations, where the quark spin flip probabilities were taken to be those of Eq.~(\ref{EQ_SPIN_FLIP}), fourth order modulations of the polarized fragmentation functions appear in simulations with just two produced hadrons in each quark decay chain. The additional two orders of modulation in $\sin(\varphi)$ originate from the quark spin flip probabilities themselves, as those depend on the squares of the components of the recoil quark transverse momentum, that can be easily expressed  in terms of $\sin^2(\varphi)$ using the transverse momentum conservation relation of the produced hadron $\vec{p}_\perp$ and the remnant quark $\vec{l}_\perp$ with respect to the fragmenting quark: $\vec{l}_\perp = -\vec{p}_\perp$ and $p_{x}^2= p_\perp^2 \cos^2(\varphi)$, $p_{y}^2= p_\perp^2 \sin^2(\varphi)$. In practice, we saw that the modulations of order higher than fourth are very hard to extract from our simulations with the present discretization of variables and the number of MC events calculated (of order $10^{11}$ in current work and $10^{12}$ in Ref.~\cite{Matevosyan:2012ga}). This is a purely computational obstacle, that can be easily solved by running the simulation on an order of magnitude more computational nodes than used for this work (only $50$ Intel Core-i7 cores ran for about two weeks), although it is hard to justify the need for such a computationally expensive exercise.

\section{Conclusions}
\label{SEC_CONCLUSIONS}
 
In this article we examined, using a toy model, the role of the quark spin flip in the quark-jet hadronization picture on the pion Collins functions  produced in Monte Carlo simulations. This work was put forward to compliment and elucidate the findings for the $1/2$ moment of the Collins function in our previous work in Ref.~\cite{Matevosyan:2012ga} and also to provide the toy model analyses needed for explanations of the higher order sine modulations in the transverse momentum dependence of the polarized fragmentation functions, first observed in the same work.

 First, the dependence of the $1/2$ moments of the Collins functions, $H_1^{\perp (1/2)}$, on the quark spin flip probability ($SF$) for simulations with only two hadron emissions was examined in Section~\ref{SEC_TOY_MODEL}. It was shown that $H_1^{\perp (1/2)}$ has a very strong dependence on $SF$, both for favored and unfavored fragmentation functions. In particular, a non-zero unfavored $H_1^{\perp (1/2)}$ is only possible if $SF\ne0.5$, with the results for $SF=0.5\pm a$, for any $0<a\leq 0.5$, having equal magnitude and opposite sign. The results with $SF=1$ have the largest magnitude and opposite sign to the input elementary favored Collins function, similar to the results of Ref.~\cite{Matevosyan:2012ga}. For the favored fragmentation functions, the results for $SF=0$ are of the same sign as the elementary favored Collins function for all values of $z$. As the $SF$ increases from $0$ to $0.5$, the low-$z$ region gets increasingly suppressed. For $SF>0.5$, they start positive (with respect to the elementary Collins function), then cross zero and become negative at small values of $z$, with minimum negative value decreasing as $SF$ increases towards $1$. The results for $SF>0.5$ again resemble those acquired in the full model calculations of Ref.~\cite{Matevosyan:2012ga}, dictating that the qualitative features of $H_1^{\perp (1/2)}$ are generated by the quark spin flip process in the hadronization process and don't depend on the particular elementary Collins functions used as input to the MC simulations.
 
 In Section~\ref{SEC_TMD} we examined the higher order modulations of the transverse momentum dependent polarized quark fragmentation functions for $SF=1$, in order to determine the role of the model spin flip probabilities of Eq.~(\ref{EQ_SPIN_FLIP}), used in Ref.~\cite{Matevosyan:2012ga}, in the mechanism that generates these modulations. We studied the order of the polynomials in $\sin(\varphi)$ needed to describe the polarized fragmentation functions from MC simulations with various numbers of the produced hadrons $N_{Links}$. We concluded that the order of these modulations is equal to $N_{Links}$ when $SF$ is kept fixed, though the numerical examinations of modulation orders larger than $4$ were limited by the variable discretization and statistical fluctuation effects. This part of the modulation was attributed to the modulation of the remnant quark in the decay with the recoil to the hadron emissions. Thus, the modulations that rise by additional two orders with a unit increase in $N_{Links}$ observed in Ref.~\cite{Matevosyan:2012ga} are attributed to the form of the model $SF$ that depends on the components of transverse momentum of the remnant quarks.
  
 \section*{Acknowledgements}
 
This work was supported by the Australian Research Council through Grants No. FL0992247 
(AWT), No. CE110001004 (CoEPP), and by the University of Adelaide. H.H.M. would like to thank the "National Center for Theoretical Sciences, Taipei, Taiwan, R.O.C." for their kind hospitality during his visit, when a part of this work has been carried out.

\bibliographystyle{apsrev}
\bibliography{../Bibliography/fragment}

\end{document}